\documentclass[
]{ceurart}

\usepackage{listings}
\begin{document}

\copyrightyear{2025}
\copyrightclause{Copyright for this paper by its authors.
  Use permitted under Creative Commons License Attribution 4.0
  International (CC BY 4.0).}

\conference{Preprint}

\title{WebXAII: an open-source web framework to study human-XAI interaction}

\author[1]{Jules Leguy}[%
orcid=0000-0002-6808-7806,
email=jules.leguy@mines-ales.fr
]
\cormark[1]
\address[1]{EuroMov Digital Health in Motion Univ Montpellier, IMT Mines Alès}

\author[1]{Pierre-Antoine Jean}[
email=pierre-antoine.jean@mines-ales.fr
]

\author{{Felipe} {Torres Figueroa}}[%
orcid=0000-0001-6747-548X,
email=felitf.94@gmail.com
]
\author[1]{Sébastien Harispe}[%
orcid=0000-0001-5630-2743,
email=sebastien.harispe@mines-ales.fr
]

\cortext[1]{Corresponding author.}

\begin{abstract}
This article introduces WebXAII, an open-source web framework designed to facilitate research on human interaction with eXplainable Artificial Intelligence (XAI) systems. The field of XAI is rapidly expanding, driven by the growing societal implications of the widespread adoption of AI (and in particular machine learning) across diverse applications. Researchers who study the interaction between humans and XAI techniques typically develop ad hoc interfaces in order to conduct their studies. These interfaces are usually not shared alongside the results of the studies, which limits their reusability and the reproducibility of experiments. In response, we design and implement WebXAII, a web-based platform that can embody full experimental protocols, meaning that it can present all aspects of the experiment to human participants and record their responses. The experimental protocols are translated into a composite architecture of generic views and modules, which offers a lot of flexibility. The architecture is defined in a structured configuration file, so that protocols can be implemented with minimal programming skills. We demonstrate that WebXAII can effectively embody relevant protocols, by reproducing the protocol of a state-of-the-art study of the literature.
\end{abstract}

\begin{keywords}
XAI \sep human evaluation \sep open-source software.
\end{keywords}

\maketitle

\section{Introduction}

AI has made tremendous strides in recent years, opening up remarkable prospects in a wide range of industrial, scientific and societal applications. This expansion raises fundamental questions about the integration of AI into socio-technical systems, particularly with regard to the governance, robustness and transparency of the models used. 
One of the critical issues lies in the ability to understand, audit and control the decisions made by complex models, particularly in the design of systems with high decision-making autonomy (e.g., autonomous agents, delegation systems).
In this context, the AI community is paying particular attention to the subject of eXplainable AI (XAI), e.g., the implementation of explicable approaches by design, 
or the proposal of approaches enabling the understanding of black-box models such as deep learning models \cite{hassija_interpreting_2024,minh_explainable_2022}. 


XAI techniques can provide insights at various stages of a model's lifecycle. They can be used as soon as the development phase, in order to debug or validate the model~\cite{bylinskii_towards_2023}. Once the model is in production, they can assist the end-user to take a decision~\cite{fok_search_2024}. They can also be used with the intent of drawing new expert domain knowledge from the model~\cite{sarcevic_cybersecurity_2022}. For sensitive domains they can also be required by law~\cite{fresz_how_2024}, or they can be needed in order to build trust in the model when the practitioner it assists is ultimately responsible of the decision~\cite{kloker_caution_2022}.

Many XAI techniques have been proposed to tackle these problems~\cite{hassija_interpreting_2024}. The current state-of-the-art tends to indicate that XAI techniques are effective in the debugging and validation phases. However, in the use case of helping an end-user take a decision based on the output of a predictive model, they have only shown mixed results yet~\cite{muller_how_2025,fok_search_2024}. Based on these results, we anticipate that numerous studies on the interaction between human participants and XAI techniques will be conducted in coming years.

The aforementioned type of human-XAI studies requires an interface to display the protocol content to participants and collect their input. In many cases, the software developed for this interface is study-specific and not reusable or shared with other researchers. In this article, we introduce WebXAII (Web XAI-Interface), an open-source web framework tailored for this purpose, i.e., providing an interface to study the interaction between human subjects and XAI techniques. WebXAII is designed to be flexible so that it can embody very different protocols, while requiring very low programming skills.

We first propose an overview of the field of XAI, and we discuss the methods related to our framework. Secondly, we describe the functionalities of WebXAII and show how it is designed to be integrated with standard protocols of the literature. Thirdly, we demonstrate it can effectively be used to reproduce the interface of a previously published study. Finally, we discuss potential developments within WebXAII which could further extend its capabilities.
\section{Related work}

To situate our contribution within the broader landscape of XAI, this section presents a short review of related work organized into three parts. We begin by summarizing the main paradigms and methods that have emerged in the field of XAI. Next, we explore how studies involving human participants interacting with XAI systems are designed. In particular, we examine taxonomies, evaluation frameworks, and general recommendations found in recent literature. Finally, we review existing software tools which can be used to facilitate the development of interfaces to conduct human-XAI interaction studies.

\subsection{XAI paradigms and methods}
XAI methodologies attempt to answer questions regarding the functioning of machine learning models. In his \textit{Mythos of Model Interpretability}, Lipton describes the reasoning behind the necessity of understanding models, but also proposes a preliminary categorization of explanation approaches, dividing the landscape of explanations between \textit{Transparency} and \textit{Post-hoc Interpretations}~\cite{mythos_interp}.

Regarding traditional machine learning models, interpretability is observed as a result of their design. This is particularly clear in models with formal definitions such as linear regression, decision trees and rule-based classifiers. In these models, the relationship between inputs and outputs is defined explicitly. One example of this relationship is the pruning of decision trees, where the removal of particular branches directly alters the prediction of the model, providing insight on the decision process. Consequently, traditional machine learning models are aligned with Lipton's Transparency, where the model's inner-workings are self-explanatory.

Most recently, Zhang et al. \cite{zhang2021survey} expanded this space with a more detailed categorization, where explanations are grouped alongside three dimensions including Lipton's earlier grouping. This categorization is divided between \textit{Active} and \textit{Passive} explanations similar to \emph{Transparency} and \emph{Post-hoc interpretations} respectively. For a second dimension, the nature of an explanation is considered, dividing them according to their type, such as \emph{attributions}, \emph{rules}, \emph{hidden semantics} and \emph{examples}. Lastly, the third dimension classifies explanations according to the scope of explanations in the input space, whether they describe attributes of individual samples or a local part of the model, or provide a global interpretation across larger portions of the dataset.

Inspired by this taxonomy, we illustrate some of the currently most widely adopted explanation methods, categorized for being attribution generated in a passive and local manner. To begin with, it is possible to use existing information within deep learning models such as the gradient computed when performing backpropagation from a given prediction \cite{simonyan2013deep,guidedbackprop,smilkov2017smoothgrad,sundararajan2017axiomatic}. In addition to the backward flow of information with gradients, Layer Relevance Propagation (\emph{LRP}) \cite{bach2015pixel} generates a representation following the computation of the relevance of key elements within the input. Most recently Concept Relevance Propagation (\emph{CRP}) \cite{Achtibat_2023} attributes this relevance to high-level concepts rather than individual features. Continuing with approaches leveraging existing information on models, Class Activation Maps (\emph{CAM}) based methods are calculated using weighted linear combinations of activation maps from intermediate layers, guided by their influence on the model’s prediction for a specific input \cite{zhou2016learning,selvaraju_grad-cam_2020,chattopadhay2018grad,wang2020score,zhang_opti-cam_2024}. Similar to CAM, Local Interpretable Model-Agnostic Explanations (\emph{LIME}) \cite{ribeiro2016should} is a local explanation method. Although LIME contains an active component, the method considers the training of surrogate models using perturbations of data within the input space to produce an explanation; nevertheless, the passive behavior is maintained since the auxiliary models do not modify the original one. Building on this idea, SHapley Additive exPlanations (\emph{SHAP}) \cite{NIPS2017_7062} also uses input perturbations but is grounded in Shapley values from cooperative game theory, ensuring fair attribution of feature importance. While local in nature, SHAP can also provide global insights by aggregating feature contributions across multiple predictions. Both SHAP and LIME are model-agnostic, meaning that they can be used with any type of machine learning model. Randomized Input Sampling for Explanation of Black Boxes (\emph{RISE})~\cite{petsiuk2018rise} is another model-agnostic method, specialized for image data. It relies on occlusion to compute a saliency map, aggregating the importance of image regions to estimate the saliency of patches within the image. Most recently, with the emergence of attention-based architectures, traditional explanation methods have shown difficulties providing meaningful explanations \cite{zhang_opti-cam_2024}; as a result the attention mechanism is used to produce explanations aggregating information across the transformer layers and heads \cite{abnar2020quantifying,chefer2021transformer,wu_token_2024}.

\subsection{Human-XAI studies design}

In the scope of our work, we are especially interested in the evaluation of XAI techniques when they are actively used in a collaborative setup with a human operator. The literature shows a growing number of articles aiming to define frameworks, or at least general principles, to standardize and improve protocols for human-XAI interaction studies.
Chromik et al. define a taxonomy of human-XAI interaction studies~\cite{chromik_taxonomy_2020}. They identify three dimensions common to every study, namely the dimension of the task, the dimension of the participants, and the dimension of the study design. Several variables are associated with each dimension and allow to characterize the study (for instance the type of task, the recruiting method or the treatment assignment, to give examples along all three dimensions).

Van der Waa et al. propose general recommendations for the design of studies~\cite{van_der_waa_evaluating_2021}. A first set of recommendations states that when evaluating an XAI method, the expected purpose of an explanation should be clearly defined as a construct, and so should the relations between the different constructs. In this way, the hypotheses related to the explanations are formed explicitly and can be tested through the experiment. A second set of recommendations is related to the use case and the experimental context. It states that they can have a very large impact on the conclusions and thus should be chosen knowingly and with care. If possible, controlled settings such as performing the experiment in person in a dedicated room should be preferred. A final set of recommendations is dedicated to measurements. It states that self-reported and behavioral measures should be used in the right context (i.e., for subjective constructs and objectively measuring constructs respectively). Ideally, both types of measures should be used to check if the reported outcome aligns with the one measured. When it comes to the effects of the explanations on the participants, the authors argue that it should be measured implicitly in order not to affect their behavior. Finally, the authors warn that bias could arise at many stages of the experiment, including during its design. It should thus be monitored closely by the experimenters.
 
Colin et al. propose an evaluation framework which is based on making the human participant a ``\textit{Meta-predictor}'', whose role is to guess the prediction of the model based on the output of the XAI technique~\cite{colin_what_2023}.
The intuitive purpose of this task is to measure how well the explanation correlates with the actual behavior of the model in a way that is informative for the human participant, while avoiding confirmation bias as only the actual performance of the human-XAI pair is evaluated. They also design a protocol based on this task to compare the relative performances of various XAI approaches.


From the literature, we shortly reviewed three articles that propose a theoretical framework, a set of study design guidelines, and a practical task integrated into an experimental protocol respectively. Although these works aim at framing and normalizing the experimental protocols, it is not clear whether their contributions will be adopted by the researchers in a near future. For now, we do not observe a clear tendency to the standardization of protocols in recent literature~\cite{muller_how_2025}.

\subsection{Developing and hosting interfaces for human-XAI studies}

Every experimental protocol which involves XAI techniques and human participants must be implemented through a software interface. Its main purposes are to give instructions to the participants, submit them to the experimental cases and collect their answers and decisions. Similarly to the protocols themselves, the interfaces which are presented in the literature follow little normalization. In addition, they are usually developed for a specific study and are only very rarely shared alongside the results of the study. This limits their reusability in other studies and may also hinder the reproducibility of results. At the very least, this makes the replication more difficult.

There does exist tools which are available to researchers and could be used to facilitate the development of such interfaces. In economics, \emph{z-Tree} has been popular for about two decades~\cite{fischbacher_z-tree_2007}. It is a free software designed for academic research (although closed source) which allows to develop interfaces for interactive studies, through a dedicated language. The interface is developed as a heavy client, which is a limitation as web interfaces are much preferred by researchers in recent years. This is because they are lightweight and enable experiments to run directly on participants' devices without installation. It should be noted that using web interfaces does require hosting the website on a server.

\emph{oTree} has been proposed as an open-source, web-based, alternative to z-Tree~\cite{chen_otreeopen-source_2016}. It provides many interaction functionalities and a more modern interface. However it does requires actual development in order to create an interface, and it does not offer specific support for human-XAI experiments as it was designed as a general-purpose tool for economic studies. 

Experiments can also be hosted online using dedicated services. The best known is probably Google Forms, which is free to use but is limited to the design of questionnaires. Thus it is too limited to handle the experimental part of the studies for human-XAI interaction and could only handle the survey phase. LimeSurvey is a more advanced commercial tool, which can be used for free if the server is self-hosted. It is typically used for surveys but it also supports more advanced features such as randomization of the questions, time limits or time recording. LimeSurvey also provides greater control over the layout and enables feedback to be given on participants’ responses. However, in such cases, it requires web development and familiarity with its custom conditional language. Qualtrics is another commercial service which provides similar features to LimeSurvey but is more expensive and is primarily intended for the user-friendly design and hosting of surveys for large organizations.

Regarding tools designed for human-XAI interaction specifically, Quispe et al. have proposed ARDAS, a \textit{micro-world} platform which emulates an industrial process, namely a hydraulic system~\cite{quispe_g_machine_2020}. It corresponds to a graphical interface which presents a schematic view of the system, historic data on the set of variables which are monitored, and actions that must be taken by the human operator. For these actions, suggestions from a machine learning model are presented, along with explanations of the suggestions through feature importance. ARDAS has been used in at least one study for human-XAI interaction~\cite{gentile_human_2023}. Since researchers often aim at studying the general performances of XAI techniques, the application to an industrial process might appear too restrictive. In addition, the source code of ARDAS is not available online, which might discourage researchers from using the platform.

Often, participants of the studies are recruited and remunerated through specialized platforms such as Amazon Mechanical Turk or Prolific. These platforms generally don't host the studies, and send the recruited participants to the external website where the experiment is hosted. 

In the following section, we introduce WebXAII, our framework for the development of interfaces for human-XAI studies. It is designed specifically for experiments which involve submitting instances of machine learning predictions and XAI explanations to the participants, and collect their choices regarding a task-specific decision. WebXAII also provides complete support of the experimental protocol, from the authentication of the participants to possible post-experiment surveys. The framework is designed to be able to embody many different protocols. The interfaces and transitions are described in a configuration file, which can be defined with very low programming expertise. WebXAII is distributed online as an open-source software, which can be easily deployed on a web server. It does not substitute to recruiting platforms but is compatible with them.
\section{The WebXAII Framework: A Functional Overview}
\label{sect_methods}

\begin{figure}
\includegraphics[width=1\textwidth]{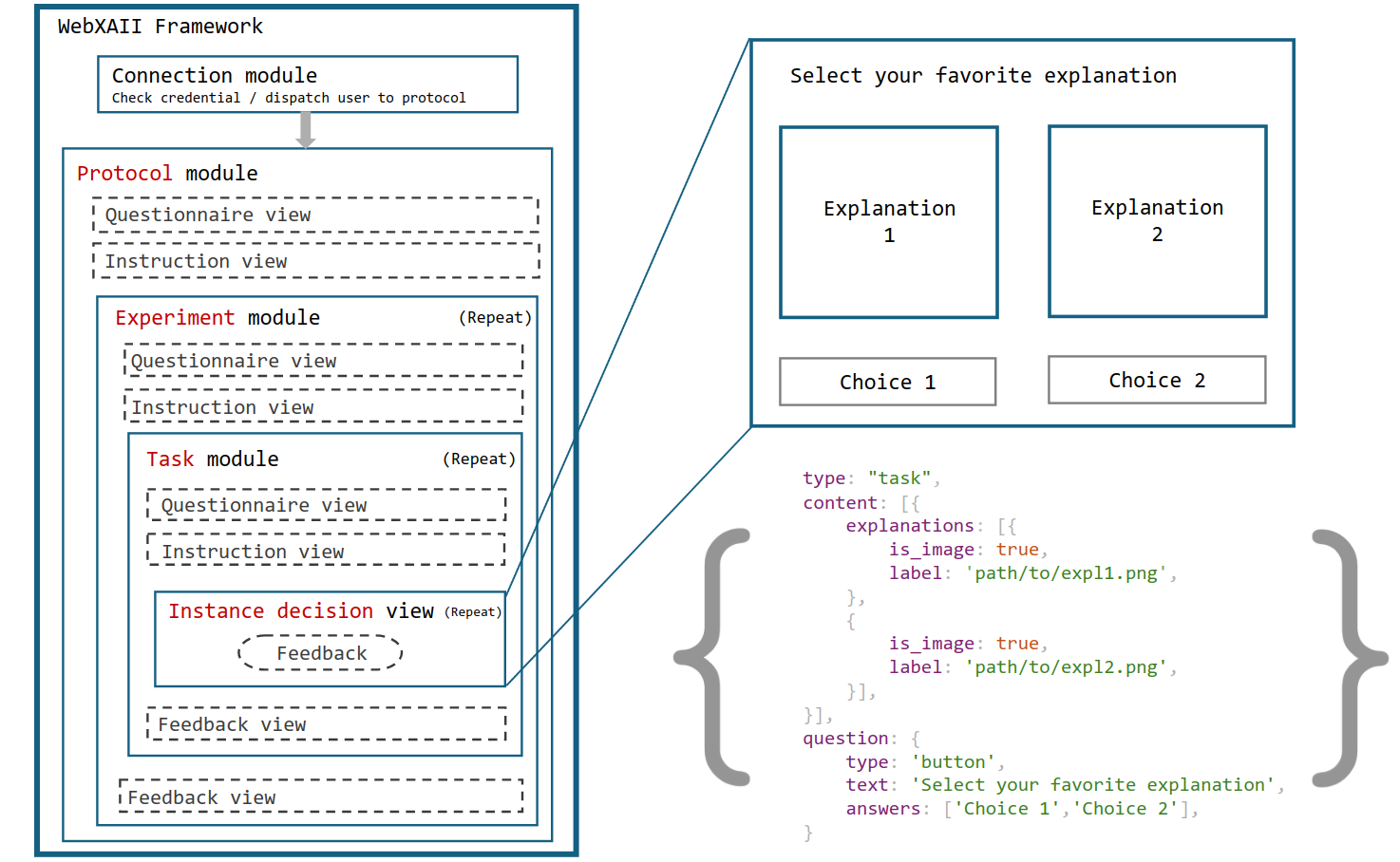}
\caption{Left : compositional representation of the main components of the WebXAII framework. The blocks outlined in dotted lines are optional. Alternative settings can be defined considering the same modules and views. Right : schematic representation of a possible Instance decision view, and JSON snippet to encode this view.} \label{fig1}
\end{figure}

WebXAII is a modular web framework developed in javascript and designed to streamline the creation and management of experimental protocols for the study of human-XAI interaction. To this end, WebXAII is built around a clear separation of concerns and a hierarchy of reusable components that are commonly used in XAI experimentations. The framework therefore provides researchers and practitioners with a flexible, standardized way to implement experimental protocols, and to collect data from participants as they perform experiments. We first discuss the general approach considered to define an experimental protocol and the main components on which it relies on. Elements regarding the functional benefits of the approach and its extensibility are discussed next.

\subsection{The WebXAII framework components}

The WebXAII framework enables full experimental setups to be defined through configurations files in JSON format. Once at least one configuration is defined, the experimental server can be deployed on a specific URL accessible to participants through the web. For simplicity, and in line with most state-of-the-art experimental settings, the framework assumes that: 
\begin{enumerate}
    \item A participant will be given a specific access to the deployed evaluation platform through a specific URL and login, i.e., we assume that the access to the platform is limited per design to avoid dealing with a complex authentication schema.
    \item The experimental protocol can be defined according to the following logical structure. A protocol contains a set of experiments, each experiment being composed of a set of tasks, and each task again being composed of a set of instance decisions (Protocol $>$ Experiment $>$ Task $>$ Instance decision). An instance decision will represent a choice made by the participant on a given instance (or sample) in the input space of the predictive model, e.g., comparing two XAI approaches on a given input image in a classification setting, as illustrated in Figure~\ref{fig1}. A task will allow to practically group a set of instances for which a common decision is submitted to the participant. This definition is deliberately broad so that it encompasses many different scenarios. An even simpler alternative to the example shown in Figure~\ref{fig1} could be to ask whether an XAI explanation generated by an XAI technique is informative or not. The experiment is meant to group a meaningful set of tasks, or to make a logical opposition between tasks that are similar but have a parameter vary. For instance, the same task as in Figure~\ref{fig1} could be repeated with a different model in a second experiment. Note that experiments are merely defined for practicality and to support the most various protocols. It is actually possible to have the whole protocol in a single experiment, which can then be transparent to the end-participant. 
    \item Computations related to AI and XAI techniques are performed prior to the implementation of the protocol. The framework will not interact with a machine learning model or with XAI algorithms, but will display their precomputed outputs through text or images.
    \item Questionnaires, Instructions and Feedback are the additional interactions we want to define with a participant. 
\end{enumerate} 

Figure \ref{fig1} presents a general diagram illustrating the framework’s composite architecture, comprising of several modules (Connection, Protocol, Experiment and Task) and of several views (Questionnaire, Instruction, Instance decision and Feedback). Each module handles a distinct stage of the participant’s journey; within each, dedicated views manage specific interactions. Considering this compositional approach, numerous experimental settings can be defined. The main components are further detailed hereafter.

\medskip

\noindent \textbf{Connection module}: this module is in charge of user management, upon arrival on the deployed WebXAII instance. Its role is to check credentials of the user, and to dispatch them onto the intended experimental protocol. Each user identifier is associated with a specific protocol beforehand. This allows running multiple protocols on the same server instance. 

\medskip

\noindent \textbf{Protocol}: the Protocol is the highest-level component, which encapsulates all the content that is submitted to a group of participants. It contains one or several experiments, which can themselves contain one or several tasks; all of which are submitted sequentially to the participants. Typically, the protocol will start with a questionnaire gathering general information about the participant, and with a set of general instructions related to the outline of experiments. If the study requires that several groups of participants are submitted to different data, this can be done by defining several protocols which will be associated with the right group of users through the connection module.

\medskip

\noindent \textbf{Experiment module}: this module encapsulates a set of tasks and a set of views, which are grouped because they form a meaningful part of the protocol. It may be repeated any number of times with potentially different settings (for multi-phase or longitudinal studies). Each iteration can include its own introductory views and a series of tasks. 

\noindent After these preparatory views, WebXAII dives into the heart of the experiment: the Task module.

\medskip

\noindent \textbf{Task module}: this module aims at grouping all components that will be used to characterize a task, i.e., a set of participant's decisions on a standardized type of question. It will contain a list of Instance decisions views (see next paragraph), each corresponding to the application of the task to a single instance. The Instance decision views can optionally be submitted in a random order to the participants. A task can typically start with an Instruction view which explains the task to the participants beforehand. It can also contain a Questionnaire view before or after the task, depending on the needs of the experiment.

\medskip

\noindent \textbf{Instance decision view}. This is the fundamental interaction point of WebXAII. It is used to define the decision context a participant will be facing during a task, e.g., evaluating an XAI technique or choosing among several choices. It can be used for any problem that consists in querying participants about the output of the model and/or the output of the XAI technique for a specific instance (or sample). Here, participants may for example be asked to examine XAI outputs - such as feature rankings, counterfactual scenarios, or decision-trees - and make choices that reflect their comprehension, trust, or preference. The Instance decision view supports exclusive or non-exclusive decisions. It is also possible to specify a time constraint for the decision making; in which case, a timer will be shown and the web application will automatically go to the next page when the time expires.

\noindent The elements related to the instance, the machine learning prediction, and the XAI techniques explanation(s) which are shown on the view are customizable depending on the needs for the task.  In the most general case, it can contain the following elements.
    \begin{itemize}
        \item One or zero representation of the instance (as an image or a text).
        \item One or zero representation of the prediction of the model for the instance (as an image or a text).
        \item Any number of XAI explanations for the instance (can be none). 
    \end{itemize}

\medskip

\noindent \textbf{Questionnaire view}. The Questionnaire view is dedicated to submitting the participant to a list of customizable questions. It can be used at all the stages of the protocol. At the start of the protocol, it can gather general information on the participant. At the end of an experiment, it can gather phase-specific opinions, attitudes toward AI, or self-assessment of instruction comprehension to name a few. Before seeing the questions related to a task, participants may also answer questions about their expectations or prior knowledge.

\medskip

\noindent \textbf{Instruction view}. Just like the Questionnaire view, the Instruction view can be used at many stages of the protocol. At the very start, it can provide general instructions related to the outline of experiments. During an experiment, it can display phase-tailored guidelines or contextualize the tasks that follow. At the end of the protocol, it can be used to inform the participants that the protocol is done, and to thank them for their participation.

\medskip 

\noindent \textbf{Feedback views}. WebXAII optionally renders a feedback to the participant at several levels. In opposition to the other views, there is not a single standardized Feedback view as the feedback can take several forms:
\begin{itemize}
\item Instance decision-level feedback: immediate confirmation or refutation of the decision with optional display of the expected answer.
\item Task progress feedback : progress bar showing the advancement of the task. This is shown as a widget in the Instance decision view.
\item Task-level feedback (scoring) : debriefing specific to the last task completed, through a score that is defined by the experimenter. It constitutes a view that is displayed any time after at least one task has been completed in the experiment.

\end{itemize}

By presenting a snippet of a configuration file, Figure \ref{fig1} also illustrates how views can be described in JSON format in a straightforward and intuitive manner.
~\\

\subsection{Functional Benefits and Extensibility}
By packaging authentication, user guidance, task definition, decision settings, and feedback into discrete, nested modules and views, WebXAII offers several key functional advantages:
\begin{itemize}
    
\item \textbf{Reusability}: standardized components can be reused across studies, reducing development time and ensuring a consistent participant experience.

\item \textbf{Configurability}: a protocol is entirely defined in a JSON file that contains the sequence and content of modules and views. This lets experimenters define full protocols without writing any code.

\item \textbf{Reproducibility} : since protocols are entirely defined through configuration files, they can easily be shared along with the result of the studies. This significantly improves the reproducibility of the experiments.

\item \textbf{Scalability}: modular repetition allows everything from a single brief survey to multi-phase, multi-task experiments to be deployed effortlessly.

\item \textbf{Traceability}: built-in data logging ensures that every questionnaire response, instruction acknowledgment and decision event is recorded.

\item \textbf{Extensibility}: WebXAII is distributed with an open-source license so that experimenters can add new features or new generic views according to their needs. We hope this will allow WebXAII to handle even more experimental protocols in the future.

\end{itemize}

\section{Case study : implementing the protocol of a study of the literature}

In this section, we implement an interface from the literature with WebXAII, to show that it can effectively be used to embody experimental protocols in the domain of human-XAI interaction. We chose to reproduce the work of Vasconcelos et al. because it presents a complete and rigorous protocol that we consider representative of the studies in the field~\cite{vasconcelos_explanations_2023}. All participants are required to perform the same tasks, but are divided into groups exposed to different experimental conditions. For each group, the protocol begins with a training phase to familiarize participants with the task. In addition, the study includes several surveys administered both before and after the experimental tasks. All these characteristics are typically found in other studies. 

The general purpose of Vasconcelos et al.'s work is to measure the impact of XAI on the overreliance of human operators towards AI systems, when humans are assisted by these systems at performing a task. Overreliance means that human tend to agree with an AI even when it is wrong, and has been observed in previous works~\cite{bucinca_trust_2021}. The article describes five studies. They are all based on a common task submitted to the participants, which is to solve a maze. More precisely, the participant must find a path in a maze which leads to an exit, and identify which is the right exit out of four propositions. The first four studies are very similar to each other, and are defined to test varying sets of hypotheses. The last study is not as typical, as participants are presented with the choice to be presented or not with explanations, which will have an impact on the difficulty of the task but also on a reward they earn when they answer correctly.

\subsection{Representing the study}

\begin{figure}
\includegraphics[width=\textwidth]{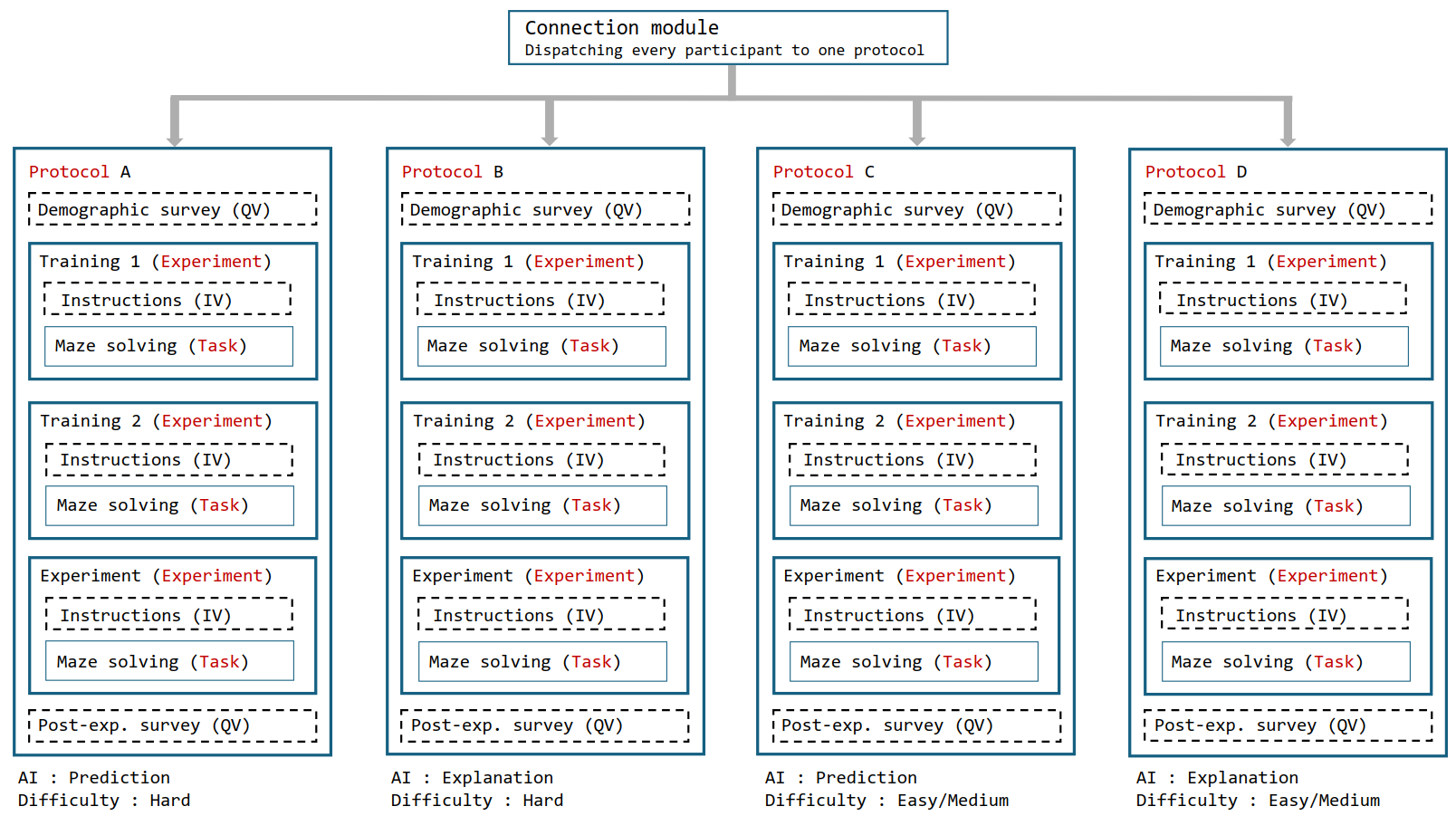}
\caption{Schematic representation of the implementation of the study with WebXAII. Each column represents a protocol. The protocols differ from each other in the conditions of the maze solving task. The conditions are specified below each protocol. QV : Questionnaire View; IV : Instructions View.} \label{fig2}
\end{figure}

For our demonstration, we decide to implement the first study of Vasconcelos et al.'s work with WebXAII. The implementation of any of the studies 2, 3 and 4 would be mostly redundant. The study 5 could currently not be implemented within WebXAII due to its specificity, which will be discussed in Section~\ref{sect_perspectives}. 

In Figure~\ref{fig2}, we represent how the study can be implemented within our framework. The outline of the study is the same for every participant, but participants are split in 4 groups which will be submitted to different conditions. In order to represent the structure of the study, we define four Protocols (as defined in Section~\ref{sect_methods}) which are named A, B, C and D. Each group of participant is directed to its corresponding protocol by the connection module.

Two variables vary through the protocols. Both of them are related to the task which is submitted to the participants. The first one is the (X)AI condition, which can take the following values:
\begin{itemize}
    \item \emph{Prediction} : the participant is presented with the maze to solve and is informed of the AI's suggestion.
    \item \emph{Explanation} : same as \emph{Prediction}, but the participant is also presented with an explanation of the suggestion. The explanation corresponds to the highlighting of the path detected by the AI in the maze.
\end{itemize}

The second one is the difficulty of the task, which depends directly on the size of the mazes. The two conditions for the study are either a mix of easy and medium mazes (\emph{Easy/Medium}), or only difficult mazes (\emph{Difficult}).

\begin{figure}
\includegraphics[width=\textwidth]{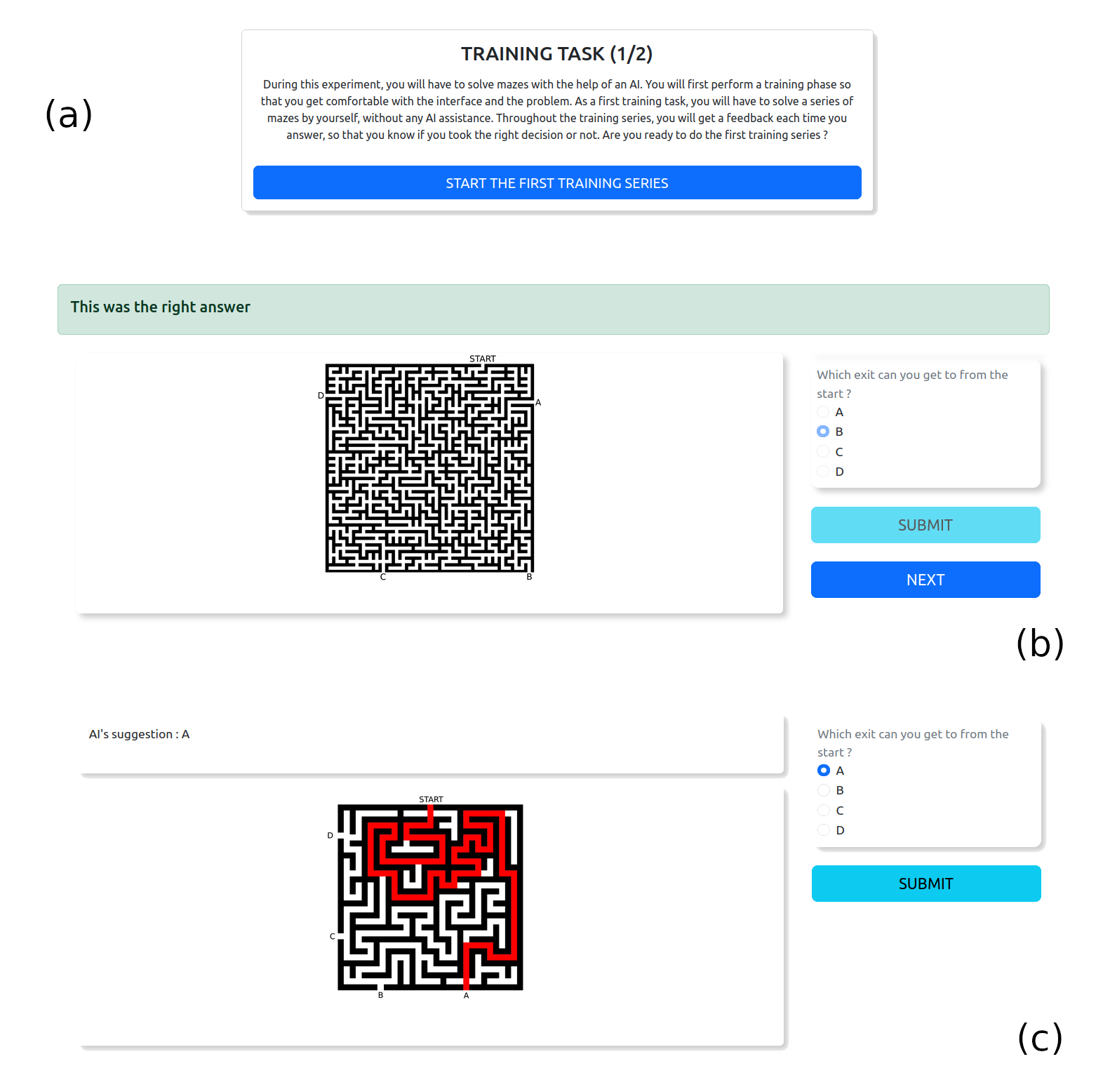}
\caption{Sample of screenshots of the interface. (a) Instruction view describing the first training experiment. (b) Instance decision view for the first training experiment (AI/XAI condition not shown) in the conditions of the protocol A or B (AI or XAI; Difficult) after submitting the decision (feedback displayed to the participant). (c) Instance decision view for the experimental phase in the conditions of the protocol D (Explanation; Easy/Medium).} \label{fig3}
\end{figure}

As illustrated in Figure~\ref{fig2}, each protocol starts with a demographic survey through a Questionnaire view. It is then followed by three experimental phases, which we represent by three Experiments (as defined in Section~\ref{sect_methods}). Each one of these experiments contains one Instruction view, and one maze-solving Task. The first two of these experiments are meant for training. In the first experiment, the participants must solve the maze by themselves, without any help from the AI. They are informed immediately if they took the right decision after each maze. The second training phase is identical but the participants also get access to the AI in the condition corresponding to their protocol. The final experiment corresponds to the actual experiment. The conditions are the same as for the second training experiment, but the participants do not get any feedback following their decisions. For all tasks, the order in which the instances (mazes) are submitted to the participants is randomized. At the end of the protocol, the participants must take a survey through a Questionnaire view.

\subsection{Implementing the study}

We implement the study within WebXAII. The implementation is meant to demonstrate that WebXAII could be used to submit the experiment to cohorts of human participants. However, we do not have access to the samples of mazes that were used in the study of Vasconcelos et al. As a consequence, we implement the structure of the study, but we only use a small sample of maze images which we generated. The JSON configuration files of the implementation are available on the git repository of the project at \url{https://github.com/PAJEAN/WebXAII}.

\begin{figure}
\includegraphics[width=\textwidth]{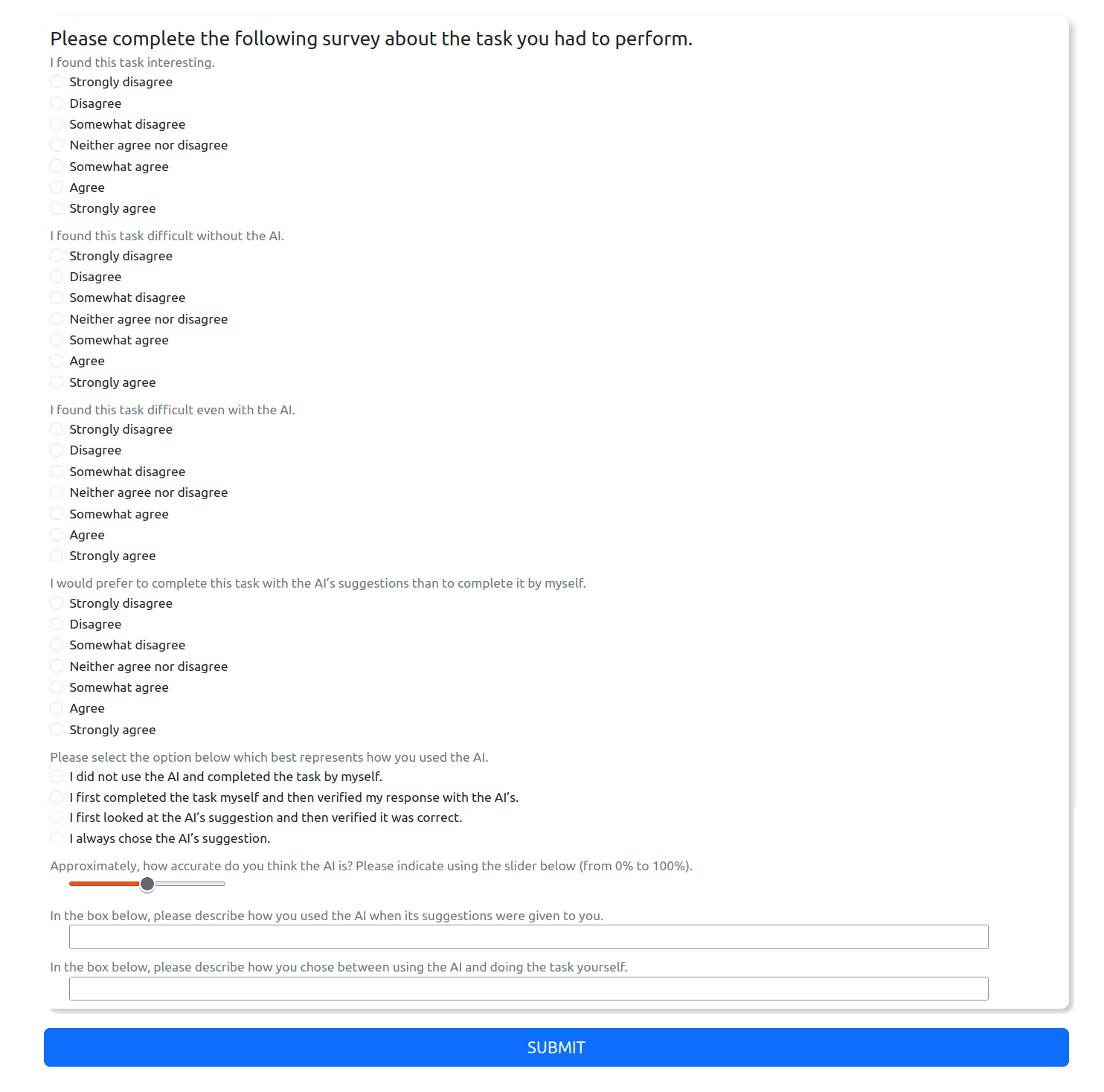}
\caption{Screenshot of the Questionnaire view for the survey at the end of the experiment.} \label{fig4}
\end{figure}

In Figure~\ref{fig3}, we present a sample of screenshots of the interface we implemented. It shows that the study can effectively be supported by WebXAII, with a clear and modern interface which provides all the features needed. The screenshot (a) shows an instance of an Instruction view, which allows to communicate instructions and information to the participant throughout the study. The Instance decision view can effectively represent the tasks which are submitted to the participants through a simple and straightforward interface. In screenshot (b), the maze it shown without any AI prediction or XAI explanation, and the feedback is given to the participant after they submitted their decision for the given instance. In screenshot (c), the maze is shown with the explanation superimposed, and the AI prediction is shown as text in the top. We remind the reader that the interface could support various combinations of elements depending on the needs of the task, as described in Section~\ref{sect_methods}. For instance, it may include multiple explanations, or the explanation(s) could be displayed alongside the sample, or the explanation or the sample could be displayed as text instead of images. Figure~\ref{fig4} presents another screenshot of the interface which shows that the Questionnaire view can collect all the feedback needed from the participants through multiple choice questions, text fields and sliders.

\section{Perspectives}
\label{sect_perspectives}

As demonstrated in previous sections, WebXAII can embody complete experimental protocols and offers extensive customization, enabling it to accommodate various protocols, tasks, and surveys. In the future, we plan to keep adding functionalities so that it supports even more use cases. For instance, regarding the Instance decision and Questionnaire views, we plan to add more features such as drag-and-drop rankings, visual toggles, select fields or text areas. 
We also plan to introduce additional options for customizing the placement of elements in the Instance decision view. New types of feedback could also be made available to experimenters. At the Experiment module level, summarized performance metrics may be supported, along with comparative insights across all tasks within the current iteration.

Currently, all interactions with machine learning and XAI techniques must be precomputed prior to the experiments. In the future, some of these interactions could be done dynamically via API calls during the application of the protocol. This would for instance allow participants to interact with a language model. Another limit is that some specific protocols could currently not be implemented within WebXAII, due to the fact that they require participants to make choices which will influence the subsequent stages of the protocol. This would for instance be the case of the last study of Vasconcelos et al.'s work~\cite{vasconcelos_explanations_2023}. Currently, protocols are defined using a list-based structure of views and modules. In the future, this could be transformed into a tree-based structure, enabling the effective implementation of such studies. Since JSON format natively supports tree-based structures, this change would have limited impact on the definition of configuration files, and could be designed to be retro-compatible with the current implementation.

Finally, we plan to implement a monitoring interface as part of the WebXAII server. This interface would help monitoring the state of the experiments run on the server instance, and would also help administrate protocols, groups of participants, etc.

\section{Conclusion}

In this article, we present WebXAII, an open-source web framework for the implementation of complete protocols for human-XAI interaction studies. It encompasses the full journey of human participants within the experiment, from login to post-experiment surveys, and including instructions, questionnaires, and experimental tasks. WebXAII is designed with a composite, view-driven architecture which makes the framework well suited to the diverse demands of XAI experimentations. Whether the goal is to compare different explanation algorithms, assess human trust in black-box models, or generally collect human responses to an XAI-related task, it provides researchers with a comprehensive framework for defining interfaces, guiding participants, collecting nuanced data, and delivering dynamic feedback. We demonstrate that it can effectively be used to reproduce the protocol of a state-of-the-art study from the literature, and we outline future directions for its development. We hope the present work will convince researchers to adopt WebXAII for their future studies.

\begin{acknowledgments}
This work is supported by the European Union’s HORIZON Research and
Innovation Programme, grant agreement No 101120657, project ENFIELD
(European Lighthouse to Manifest Trustworthy and Green AI).
\end{acknowledgments}

\section*{Declaration on Generative AI}
During the preparation of this work, the author(s) used ChatGPT in order to: Paraphrase and reword. After using these tool(s)/service(s), the author(s) reviewed and edited the content as needed and take(s) full responsibility for the publication’s content. 

\section*{Contributions}
JL lead the project and is the primary writer of the article. PAJ developed WebXAII. FTF helped reproducing the interfaces of the literature and participated to the redaction. SH supervised the work and participated to the redaction. All authors read and approved the final manuscript.

\section*{Availability of code}
WebXAII is fully available at \url{https://github.com/PAJEAN/WebXAII}.

\bibliography{biblio}

\end{document}